\documentclass[preprint,showpacs,preprintnumbers,amsmath,amssymb,nofootinbib]{revtex4}
\usepackage{graphicx}
\usepackage{dcolumn}
\usepackage{bm}

\usepackage[latin1]{inputenc}
\usepackage{shortvrb}
\usepackage{amssymb}
\usepackage{amsmath}
\usepackage{amssymb}
\usepackage[usenames]{color}

\newcommand{\beq}{\begin{equation}}
\newcommand{\eeq}{\vspace{0cm} \end{equation}}
\newcommand{\beqq}{\setlength\arraycolsep{2pt}\begin{eqnarray}}
\newcommand{\eeqq}{\vspace{0cm} \end{eqnarray}}

\newcommand{\h}{\hbar}

\begin{document}

\title{On the Sackur-Tetrode equation in an expanding universe}

\author{S. H. Pereira} \email{saulopereira@unifei.edu.br}

\affiliation{Universidade Federal de Itajub\'a, Campus Itabira \\ Rua S\~ao Paulo, 377 -- 35900-373, Itabira, MG, Brazil}

\bigskip

\begin{abstract}
In this work we investigate the thermodynamic properties satisfied by an expanding universe filled with a monoatomic ideal gas. We show that the equations for the energy density, entropy density and chemical potential remain the same of an ideal gas confined to a constant volume $V$. In particular the Sackur-Tetrode equation for the entropy of the ideal gas is also valid in the case of an expanding universe, provided that the constant value that represents the current entropy of the universe is appropriately chosen.


\vspace{1cm}

En el presente trabajo investigamos las propriedades termodinámicas que son satisfechas por un universo en expansión, el cual es lleno por un gas ideal monoatómico. Se prueba que las ecuaciones para la densidad de la energía, la densidad de la entropía y el potencial químico son las mismas que las de un gas ideal, el cual se encuentra confinado en un volumen V.  En particular, la ecuación de Sackur-Tetrode, para la entropía del gas ideal continua siendo válido en el caso de un universo en expansión, siempre que el valor constante que representa la entropía del universo actual sea escojido adecuadamente.

{\it Descriptores}: Universo en expansión; gas ideal; ecuación de Sackur-Tetrode.
\end{abstract}


\maketitle

\section*{INTRODUCTION}

The thermodynamic properties of an ideal gas have been known since the beginning of the nineteenth century through the works of Clapeyron, Boltzmann and other. Although it is only a theoretical model, since the particles are considered isolatedly and not interacting with each other, the ideal gas is a model that reproduces with great precision most of the gases at high temperature and low pressure. At normal conditions such as standard temperature and pressure, most real gases behave qualitatively like an ideal gas. Many gases such as air, nitrogen, oxygen, hydrogen, noble gases, and some heavier gases like carbon dioxide can be treated like ideal gases within reasonable tolerances \cite{gi01}. The theoretical model of the ideal gas tends to fail to describe substances at lower temperatures or higher pressures, when intermolecular forces and molecular size become important. It also fails for most heavy gases, such as water vapor \cite{gi01}. These must be modeled by more complex equations of state.

Recently the model of the ideal gas has been object of several studies, both in journals of education and research. Among the studies related to education we can quote the spectral and thermodynamical properties of systems with noncanonical commutation rules \cite{gi02}, the exact calculation of the number of degrees of freedom of
a rigid body composed of $n$ particles \cite{gi03}, the theoretical aspects concerning the thermodynamics of an ideal bosonic gas trapped by a harmonic potential \cite{gi04} and the description of the ideal gas free expansion obtained with the aid of a computational modeling \cite{gi05}. We can also cite numerous research topics, including cosmological applications \cite{gi06}, interaction with a nonrelativistic Kaluza-Klein gas \cite{gi07}, statistical properties of a two dimensional relativistic gas \cite{gi08} and Einstein's equations for spherically symmetric static configurations of ideal gas \cite{gi09}. All this shows that the ideal gas model is still being widely exploited and sometimes generalized to many different applications.

The relationship between the internal energy $ U $ of an ideal gas and the pressure $ P $ is given by the equation of state $ PV =  {2\over 3} U$. This relationship is very close to that of a very important real system, the electromagnetic radiation, which satisfies $PV={1\over 3}U$.

Recently there has been a great interest in the study of thermodynamic systems  satisfying an equation of state in the general form $PV=\omega U $, with $\omega$ a constant, positive or negative. The main motivation is that several kinds of complementary astronomical observations indicate
that the Universe is expanding in an accelerated manner \cite{ref1}.  In the context
of general relativity, an accelerating stage and the associated
dimming of type Ia Supernovae  are usually explained by assuming the
existence of an exotic substance with negative pressure sometimes
called dark energy, with  $\omega < -{1\over 3}$. (See \cite{ref2} for a good review and \cite{saulo1} for a recent discussion including chemical potential.)

There are many candidates to represent this extra non-luminous
relativistic component. In the case of cold dark matter cosmologies, for
instance, it can be phenomenologically  described by an  equation of
state of the form $PV=\omega U $.  The case $\omega = -1$ corresponds to a positive cosmological constant, or vacuum energy, while for $\omega < -1$ we
have the so called phantom dark energy regime \cite{phantom}, or phantom 
fluids. Indeed, in the standard lines of the thermodynamic, 
we have showed in \cite{saulo2} that it does not make sense to speak of phantom fluids for systems with null chemical potential.

In this article we consider only the case $\omega = 2/3$, which corresponds to a universe filled with a monatomic ideal gas. We will show that the evolution laws for energy, entropy and chemical potential are reduced to those of an ideal gas confined to a constant volume V, even though the laws of evolution seem to be quite different in an expanding universe.

In section 2 we review the thermodynamic properties of a monoatomic ideal gas. In section 3 we present the thermodynamics of an expanding universe with an equation of state in the general form. In section 4 we consider the particular case of a universe filled with a monoatomic ideal gas, and we show that the same properties of the second section can be obtained. We finish by presenting a simple estimate for the entropy and chemical potential of a neutrino gas filling the universe, showing that this estimate is in accordance with the limits set by primordial nucleosynthesis theory.

\section{The monoatomic ideal gas}

Considering the classical thermodynamics, it is well known that the entropy of a classical ideal gas can be given only within a constant. For a monoatomic classical ideal gas, an exact expression can be reached using quantum considerations. At the beginning of the last century, around 1912, Hugo Tetrode and Otto Sackur independently developed  an equation for the entropy using a solution of the Boltzmann statistic. This equation is named Sackur-Tetrode equation, and is represented by \cite{termo}
\beq
S={5\over 2}k_BN + k_BN\ln\bigg({V\over N}\bigg)+{3\over 2}k_BN\ln\bigg({mk_BT\over 2\pi\hbar^2}\bigg)\,,\label{e1}
\eeq
where $k_B = 1.381\times 10^{-23}$J/K is the Boltzmann constant, $N$ is the particle number, $V$ is the volume, $T$ the temperature of the gas, $m$ is the mass of the particle and $\h = 1.054 \times 10^{-34}$J.s is the Planck constant. The last term represents the quantum correction. In classical thermodynamics this constant remains undefined, and it can be determined only through the quantum statistical treatment. This expression can be reduced to a more compact form
\beq
S=Nk_B\ln \bigg[\exp(5/2){V\over N}\bigg({mk_BT\over 2\pi\h^2}\bigg)^{3/2}\bigg]\,.\label{e1a}
\eeq

The monoatomic ideal gas satisfies the ideal gas law $PV=Nk_BT$ and its internal energy is given by
\beq
U={3\over 2}Nk_BT\,.\label{e2}
\eeq
The pressure $P$ is related to the energy $U$ by the equation of state
\beq
PV={2\over 3}U\,.\label{e3}
\eeq
Finally, the chemical potential of the ideal monoatomic gas can be obtained by
\beq
\mu=\bigg({\partial G\over \partial N}\bigg)_{T,P}\,,\label{e4}
\eeq
 where $G=U+PV-TS$ is the Gibbs free energy. By substituting the above expressions we obtain
\beq
\mu = -k_BT\ln\bigg[{V\over N}\bigg({mk_BT \over 2\pi \h^2}\bigg)^{3/2}\bigg]\,.\label{e5}
\eeq
An interesting aspect about this equation is the negative sign. Why is it negative? This question was very well explored by Cook and Dickerson \cite{cook}: ``Actually, in the classical limit, the quantity in square brackets is large, much greater than 1, making $\mu$ a negative number. This is so whenever  $T$ is large, and the volume per particle, $V/N$, is large compared to the cube of the thermal de Broglie wavelength $\lambda = h/\sqrt{3mk_BT}$. In fact, $\mu$ must be negative, because in order to add a particle, while keeping the entropy and volume constant, the particle must carry negative energy, or rather, it must be added while the internal energy of the ideal gas is allowed to decrease, by cooling''. In the last section we make a simple application to a gas of neutrinos and we see that in fact the chemical potential is negative in our simplified model.

Followings we will consider that the volume varies, so it is more convenient to express the above equations in terms of the energy density $\rho\equiv U/V$, the particle number density $n\equiv N/V$ and the entropy density $s\equiv S/V$,
\beq
P={2\over 3}\rho\,,\label{e6}
\eeq
\beq
\rho={3\over 2}nk_BT\,,\label{e7}
\eeq
\beq
s=nk_B\ln \bigg[{\exp(5/2)\over n}\bigg({mk_BT\over 2\pi\h^2}\bigg)^{3/2}\bigg]\,,\label{e8}
\eeq
\beq
\mu = -k_BT\ln\bigg[{1\over n}\bigg({mk_BT\over 2\pi\h^2}\bigg)^{3/2}\bigg]\,.\label{e9}
\eeq

Our aim is to show that these equations remain valid even in an expanding universe where the thermodynamical parameters are not constant.

\section{Thermodynamics of an expanding universe}

Let us consider that the universe is described by the homogeneous and isotropic Friedmann-Robertson-Walker geometry \cite{frw,kolb} and is filled with a fluid described by the general equation of state
\beq
P=\omega \rho\,,\label{e10}
\eeq
where $\omega$ is a constant parameter.

The equilibrium thermodynamic states of a relativistic simple
fluid obeying the equation of state (\ref{e10}) can be completely characterized
by the conservation laws of energy, the number of particles, and entropy.
In terms of specific variables $\rho$, $n$ and $s$, the conservation laws can be expressed as
\begin{equation}
 \dot{\rho} + 3 (1 + \omega)\rho \frac{\dot a}{a}=0,\,\,\;\;
 \dot{n} + 3n\frac{\dot a}{a}=0,\,\,\;\; \dot{s} + 3s\frac{\dot a}{a}=0, \label{e11}
\end{equation}
where $a\equiv a(t)$ is the scale factor of the evolution, or roughly speaking, the universe radius, so that $V\propto a^3$ varies with the universe expansion. The above equations have general solutions of the form:
\begin{equation}
\rho=\rho_0 \left(\frac{a_0}{a} \right)^{3(1 + \omega)}, \,\, n=n_0
\left(\frac{a_0}{a}\right)^{3}, \,\, s=s_0
\left(\frac{a_0}{a}\right)^{3},\label{eqSol}
\end{equation}
where $\rho_0$, $n_0$, $s_0$ and $a_0$ are present day (positive) values of the
corresponding quantities. On the other hand, the
quantities $P$, $\rho$, $n$ and $s$ are related to the temperature
$T$ by the Gibbs law
\begin{equation} \label{eq:GIBBS}
nTd\big({s\over n}\big)= d\rho - {\rho + p \over n}dn,
\end{equation}
and from the Gibbs-Duhem relation there are only two
independent thermodynamic variables, say $n$ and $T$. Therefore,
by assuming that $\rho=\rho(T,n)$ and $P=P(T,n)$, one may show that the
following thermodynamic identity must be satisfied 
\begin{equation}
T \biggl({\partial P \over \partial T}\biggr)_{n}=\rho + P - n
\biggl({\partial \rho \over \partial n}\biggr)_{T},
\end{equation}
an expression that remains locally valid even for out of equilibrium states \cite{weinb}. 
Now, inserting the above expression into the energy conservation law as
given by (\ref{e11}) one may show that the temperature satisfies
\begin{equation} \label{eq:EVOLT}
{\dot T \over T} = \biggl({\partial P \over \partial
\rho}\biggr)_{n} {\dot n \over n} = -3\omega \frac{\dot a}{a},
\end{equation}
and assuming $\omega \neq 0$ a straightforward integration yields
\begin{equation} \label{eq:TV}
T=T_0 \left(\frac{a}{a_0} \right)^{-3\omega}\,,
\end{equation}
so that the equations (\ref{eqSol}) can be written in terms of the temperature as
\begin{equation}
\rho=\rho_0\left(\frac{T}{T_0}\right)^{(1+\omega)/\omega}\,,\,\,\; s=s_0\left(\frac{T}{T_0}\right)^{1/\omega}\,,\,\,\;
n=n_0\left(\frac{T}{T_0}\right)^{1/\omega}\,.
\label{resultados}
\end{equation}
These relations simply tell us that today, when the temperature of the universe is $T_0$, these quantities are equal to $\rho_0$, $n_0$ and $s_0$, which represent constants still undefined. Let us talk a little about these constants, particularly $s_0$. Just as the classical ideal gas entropy is defined up to an additive constant, here we find the same problem because we can not calculate the current value of the entropy of the universe, neither the number of particles $n_0$ and energy densities $\rho_0$. We need some model to infer the values of these constants. The observational data and recent theoretical models constrain very accurately the values of these constants. In the last section we will analyze a simple model of a gas of neutrinos and show that our estimate is in agreement with one of these models.

Finally, let us see how the expression for the chemical potential is. Using the relation (\ref{e4}), the chemical potential is given by
\beq\label{muf}
\mu=-\bigg[{ s_0T_0 - (1+\omega)\rho_0 \over n_0}\bigg]{T\over T_0}\,.
\eeq
Note that the negative sign was purposely left in evidence.

\section{Universe filled with a monoatomic ideal gas}

In order to shown that the equations of a monoatomic ideal gas remain valid even in an expanding universe, let us take $\omega=2/3$ in the above equations:
\begin{equation}
\rho=\rho_0\left(\frac{T}{T_0}\right)^{5/2},\,\,\; n=n_0\left(\frac{T}{T_0}\right)^{3/2},\,\,\;
s=s_0\left(\frac{T}{T_0}\right)^{3/2},\,\,\;\mu=-\bigg[{s_0 T_0\over n_0} -{5\over 3} {\rho_0\over n_0}\bigg]{T\over T_0}.\label{e13}
\end{equation}
A very interesting feature of these equations is that all of them have a temperature dependence very different from those of equations (\ref{e7})-(\ref{e9}). We also see that as the universe is cooling down, all these quantities decrease with evolution.

Let us consider the equation for the energy density. It can be written as
\beq
\rho=\rho_0\bigg({T\over T_0}\bigg)^{5/2}=\rho_0\bigg({T\over T_0}\bigg)\bigg({T\over T_0}\bigg)^{3/2}=\rho_0{n\over n_0}\bigg({T\over T_0}\bigg)={\rho_0\over n_0 T_0}nT\,.
\eeq
But this expression has exactly the same form as Eq. (\ref{e7}) if we define
\beq
{\rho_0\over n_0 T_0}\equiv {3\over 2}k_B\,,\label{e131}
\eeq
so that
\beq
\rho={3\over 2}nk_BT\,.\label{e71}
\eeq
Thus we see that the energy density of the expanding universe behaves exactly like that of an ideal gas. Furthermore, the expression (\ref{e131}) is very interesting, relating the present day values of the constants $n_0,\, \rho_0$ and $T_0$ with the Boltzmann constant $k_B$.

Regarding the entropy expression, apparently the Sackur-Tetrode equation (\ref{e8}) has nothing to do with the corresponding one of equation (\ref{e13}). In the first, the temperature dependence is logarithm, while in the second, it is a power law. But note that $ s_0 $ is a constant that needs to be determined, corresponding to the actual entropy of the universe. Defining the $s_0$ constant as
\beq
s_0=k_Bn_0\ln\bigg[{\exp(5/2)\over n_0}\bigg({T_0 m k_B\over 2\pi \h^2}\bigg)^{3/2}\bigg]\label{e132}
\eeq
and substituting in the entropy expression we have, after some algebraic manipulations
\beqq
s&=&s_0\bigg({T\over T_0}\bigg)^{3/2}\nonumber\\
&=&s_0{n\over n_0}\nonumber\\
&=&nk_B\ln\bigg[{\exp(5/2)\over n_0}\bigg({T_0 m k_B\over 2\pi \h^2}\bigg)^{3/2}\bigg]\nonumber\\
&=&nk_B\ln\bigg[\exp(5/2)\bigg({m k_B T\over 2\pi \h^2}\bigg)^{3/2}{1\over n_0}\bigg({T_0\over T}\bigg)^{3/2} \bigg]\nonumber\\
&=&nk_B\ln\bigg[{\exp(5/2)\over n}\bigg({ m k_B T\over 2\pi \h^2}\bigg)^{3/2}\bigg]\,,\label{entro}
\eeqq
which is exactly the Sackur-Tetrode equation.

Finally, we analyze the chemical potential. Using the relations (\ref{e131}) and (\ref{e132}) and substituting,
\beqq
\mu &=&-\bigg[{s_0\over n_0} - {5\over 3}{\rho_0\over n_0 T_0}\bigg]T \nonumber \\
&=& -k_BT\left[\left(\frac{s_0}{k_Bn_0}-{5\over 2}\right)\right]\nonumber \\
&=&-k_BT\Bigg[\ln\bigg[{\exp(5/2) \over n_0}{\bigg({T_0mk_B\over 2\pi \hbar^2}\bigg)^{3/2}}-\ln[ \exp(5/2)]\bigg]\Bigg]\nonumber \\
&=&-k_BT \ln\left[{1\over n_0}\left(\frac{mk_BT_0}{2\pi\hbar^2}\right)^{3/2}\right]\nonumber\\
&=&-k_BT \ln\left[{1\over n}\left(\frac{mk_BT}{2\pi\hbar^2}\right)^{3/2}\right]\,,
\eeqq
exactly the same as obtained in (\ref{e9}).

Thus we show that all the equations of an ideal gas are still valid even in the case of an expanding universe.

\section{Concluding remarks}

We studied the thermodynamic properties satisfied by an expanding universe filled with a monoatomic ideal gas. We showed that, although the relationships appear to be different, when we define  the constants $\rho_0 $ and $s_0$ in a convenient way, the same relations of an ideal gas confined in a region of constant volume $V$ are obtained.

In order to verify the above equations, we make some simple estimates with this model assuming that neutrinos can be represented by the equation of state of an ideal gas, satisfying $ P= 2/3\rho $. This is a very rough approximation, but we will see that the results are in agreement with more sophisticated models. Neutrino cosmology is a very current topic of research in astrophysics \cite{frw,kolb}. The estimate for the current density of neutrino is of order $n_{\nu 0}=115.05\times 10^6$ particles/m$^3$, which corresponds to $n_{0}=8.84\times 10^{-40}$GeV$^3$ in natural units \cite{units}, and the neutrino energy density is $\rho_{0}=2.95\times 10^{-15}$J/m$^3$ $= 1.41\times 10^{-51}$GeV$^4$. One of the successful predictions of primordial nucleosynthesis provide strong indirect evidence that for each neutrino flavor the relation (see page 180 of \cite{kolb})
\beq
{n \over s} < 1\label{neut}
\eeq
is satisfied. The primordial nucleosynthesis also sets an upper limit to the individual neutrino chemical potentials of \cite{kolb}
\beq
\mu < 8.9\times 10^{-3}\textrm{eV}\,\,.\label{chem}
\eeq

Using the model of an ideal gas, we can estimate the entropy density $s_0$ of a neutrino gas filling the whole universe and then verify if this relation is satisfied today, when the temperature $T_0$ is of the order of the temperature of the cosmic microwave background , $T_0=2.75$K. Taking the neutrino mass as $m_\nu = 0.01$eV$=1.78\times 10^{-38}$kg (a reasonable value), and substituting it into the equation (\ref{e132}), we get $s_{0}=1.28\times 10^{-14}$J/Km$^3$ $=9.29\times 10^{8}$/m$^3$. Thus we see that the relation (\ref{neut}) is satisfied, since we obtain ${n_{0}/ s_{0}=0.12}$ in natural units.

Another important result is the calculation for the chemical potential of the neutrino gas filling the universe. Making the substitutions in the equation (\ref{e13}), we get $\mu = -1.76\times 10^{-22}$J $=-1.10\times 10^{-3}$eV for the current value of the chemical potential. We see that the relation (\ref{chem}) is satisfied and the negative sign shows that it is in agreement with what was previously discussed in section 1. 

Remember that the ideal gas model is valid for high temperatures, but in this work we are making estimates using a very low value to $T_0$, which is the temperature of cosmic background radiation of $2.75$K, and even so we get a good agreement with the predictions of other models. This shows that the ideal gas model is a powerful tool to make estimates and even obtain good results for some situations that apparently require more sophisticated models.

\begin{acknowledgements}
I am grateful to Juan C. Z. Aguilar for reading the manuscript and  to CNPq - Conselho Nacional de Desenvolvimento Científico e Tecnológico, brazilian research agency, for the financial support, process number 477872/2010-7.
\end{acknowledgements}


\begin{thebibliography}{99}

\bibitem{gi01} Y. A. Cengel, M. A. Boles and A. Michael, {\it Thermodynamics: An Engineering Approach}, 4th Edition, Mcgraw-Hill, (2001).

\bibitem{gi02} J. C. Flores and S. Montecinos, Rev. Mex. Fis. 50(3) 221 (2004).

\bibitem{gi03} J. Bernal, R. Flowers-Cano, and A. Carbajal-Dominguez, Rev. Mex. Fis. 55 (2) 191 (2009).

\bibitem{gi04} V. Romero-Roch\'in and V. S. Bagnato, Braz. J. Phys. 35, 607 (2005).

\bibitem{gi05} L. M. Amador and J. F. Betancourt, Braz. J. Phys. 27, 428 (2005).

\bibitem{gi06}  R. A. Sussman and J. Triginer, Class. Quant. Grav. 16, 167 (1999); Ying-Qiu Gu, arXiv:0708.2962v3 [physics.gen-ph] (2009).

\bibitem{gi07} J. L. F. Barbon and C. A. Fuertes, Phys. Rev. D80 026006, (2009).  

\bibitem{gi08} A. Montakhab, M. Ghodrat and M. Barati, Phys. Rev. E79, 031124 (2009).

\bibitem{gi09} J. Gersl, Int. J. Mod. Phys. D18, 763, (2009). 

\bibitem{ref1} A. G. Riess {\it et al.}, Astron. J. {\bf{116}}, 1009 (1998);S. Perlmutter {\it et al.}, Astrophys.  J. {\bf{517}}, 565 (1999);  P. Astier et al., Astron. Astrophys. {\bf 447}, 31 (2006); A. G. Riess et al., Astro. J. {\bf 659}, 98 (2007).

\bibitem{ref2} T. Padmanabhan, Phys. Rept. {\bf 380}, 235 (2003); P. J. E. Peebles and  B. Ratra,  Rev. Mod. Phys. {\bf 75}, 559 (2003);
J. A. S. Lima, Braz. J. Phys. {\bf 34}, 194 (2004); J. S. Alcaniz, Braz. J. Phys. {\bf 36}, 1109 (2006); V. Sahni and A. Starobinsky, IJMP {\bf D15},
2105 (2006).


\bibitem{saulo1} J. A. S. Lima and S. H. Pereira, Phys. Rev. {\bf D78}, 083504 (2008).

\bibitem{phantom} R. R. Caldwell, Phys. Lett. B {\bf 545}, 23 (2002); S. M. Carroll, M.
Hoffman and M. Trodden, Phys. Rev. D {\bf 68}, 023509 (2003); B.
McInnes, [astro-ph/0210321]; V. Faraoni, Int. J. Mod. Phys. D {\bf 11},
471 (2002);  P. F. Gonzalez-Diaz, Phys. Rev. D {\bf 68}, 021303
(2003); Y.S. Piao and E. Zhou,
Phys. Rev. D {\bf 68}, 083515 (2003); M. Sami and A. Toporensky,
[gr-qc/0312009]; J. M. Cline, S. Jeon, G. D. Moore, [hep-ph/0311312]; R.
Silva, J. S. Alcaniz and J. A. S. Lima, Int. J. Mod. Phys. D {\bf
16}, 469 (2007).

\bibitem{saulo2} S. H. Pereira and J. A. S. Lima, Phys. Lett. {\bf B669}, 266 (2008).

\bibitem{termo} S\'ilvio R. A. Salinas, {\it Introduc\~ao \`a F\'isica Estat\'istica}, EDUSP, S\~ao Paulo, Brasil, (1997); H. B. Callen, {\it Thermodynamics and an Introduction to Thermostatistics}, 2nd Edition, John Willey, New York (1985).

\bibitem{cook} G. Cook and R. H. Dickerson, Am. J. Phys. {\bf 63} (8), 737 (1995).

\bibitem{frw} S. Weinberg, {\it Cosmology}, Oxford University Press, New York,
USA (2008); E. W. Kolb and M. S. Turner, {\it The Early Universe}, Westview Press, USA, (1990); V. Mukhanov, {\it Physical Foundations of Cosmology}, Cambridge University Press, (2005).


\bibitem{kolb}E. W. Kolb and M. S. Turner, {\it The Early Universe}, Westview Press, USA, (1990);

\bibitem{weinb} S. Weinberg, Astrop. J. {\bf 168}, 175 (1971); R.
Silva, J. A. S. Lima and M. O. Calv\~ao, Gen. Rel. Grav. {\bf 34},
865 (2002).

\bibitem{units} In natural units the following conversion factors are valid (see page 499 of \cite{kolb} for more conversion factors): 1 GeV $=1.7827\times 10^{-27}$kg;  1 GeV$^{-1}$ $=1.9733\times 10^{-16}$m;  1 GeV $=1.1605\times 10^{13}$K.

\end{thebibliography}
\end{document}